\documentstyle[epsf,preprint,floats,tighten,aps]{revtex}
\begin{document}
\title{First-principles calculations for Fe impurities in KNbO$_3$}
\author{A.~V.~Postnikov$\dagger$, A.~I.~Poteryaev$\ddagger$ and
        G.~Borstel$\dagger$}
\address{$\dagger$University of Osnabr\"uck -- Fachbereich Physik,
         D-49069 Osnabr\"uck, Germany\\
	 $\ddagger$Institute of Metal Physics,
	 Academy of Sciences of Russia, Yekaterinburg GSP-170, Russia}
\maketitle
\begin{abstract}
Nb-substituting Fe impurity in KNbO$_3$ is studied
in first-principles supercell calculations
by the linear muffin-tin orbital method.
Possible ways to account for the impurity charge
compensation are discussed. Calculations
are done in the local density approximation (LDA) and,
for better description of Coulomb correlation effects
within the localized impurity states, also in the LDA+$U$
scheme. The achievements and problems encountered
in both approaches are analyzed.
It is found that the impurity possess either a low-spin
configuration (with 0 or 1 compensating electron),
or a high-spin configuration (with 2 or 3 compensating
electrons), the latter two apparently corresponding to
practically relevant rechargeable impurity states.
\end{abstract}

\section{Introduction}
\label{sec:intro}

Impurities in ferroelectric crystals play an important r\^{o}le
in many physical applications, particularly in what regards
optical absorption and the change of dielectric properties
under illumination, the so-called photorefractive effects.
For a review on this subject see, e.g., Ref.~\cite{review}.
The explanation of the photorefractive effect
is essentially based on the assumption of the existence
of several charge state of an impurity, which can be
switched by the drift of electrons or holes in the process
of illumination. In what regards specifically Fe-doped KNbO$_3$,
one can address Ref.~\cite{buse95} for the description
of the models available, and Ref.~\cite{medra} for practical
aspects of tuning photorefractive properties.
Therefore, much experimental effort has been
concentrated on the elucidating the electron configuration
of impurities in question, related lattice relaxation,
or other defects coming along with the substitutional impurities.
Important information of this kind can be extracted from
electron paramagnetic resonance measurements. Earlier studies
for Fe in KNbO$_3$\cite{siegel} established the presence
of Fe$^{3+}$ centers without local charge compensation
and also of other centers, that were interpreted as
Fe$^{3+}$ with local charge compensation, presumably
Fe at Nb site with an oxygen vacancy in the nearest neighborhood.
The subsequent studies\cite{donner} included more detailed
comparison of the spectra with the description based on different
structure distortion models. It was suggested that
the Fe$^{3+}$ ion is displaced by $\sim$0.2{\AA} towards
the neighboring O vacancy. There seems to be now general agreement
between the experimentalists that Fe preferentially enters
the Nb site in KNbO$_3$. The calculations of the energy balance
with empirical interaction potential (see Ref.~\onlinecite{exner},
with the potentials from Ref.~\onlinecite{lewis}) show beyond
doubt that the Fe substitution at the K site is energetically
unfavorable. In contrary to this, Ref.~\cite{beck} claimed
that Fe mostly substitutes K in KNbO$_3$, based on the similarity
of angular yield particle-induced X-ray emission profiles
for K and Fe. This similarity however does not seem highly
convincing. One should note that the Fe doping in KTaO$_3$,
that is equistructural and equielectronic to KNbO$_3$, favors another
pattern, with higher probability of entering the K site\cite{exner}.

In the situation when the information on the electronic structure
of doped systems is available from experiment in a rather
indirect way, there is a demand for {\it ab initio} or other
reliable calculations which would provide the energy positions
of impurity levels in the gap and to compare energetically
different mechanisms for the charge compensation and the
relaxation pattern around impurity. To out knowledge, just two
attempts have been undertaken in this direction up to now.
The result of a shell-model calculation by Exner\cite{exner}
for a Fe at Nb site and an oxygen vacancy as its nearest neighbor
is that Fe is displaced by 0.38{\AA} {\it from} the vacancy, i.e.,
contrary to what was supposed based on the analysis of
the electron spin resonance data\cite{donner}. Also, the
displacement of four ``equatorial'' O neighbors to Fe
is 0.19{\AA} outwards in the shell-model calculation\cite{exner},
contrary to the estimated 0.1{\AA} inwards in Ref.~\cite{donner}.
The {\it ab initio} calculation with the linearized
muffin-tin orbital method (LMTO) in the atomic sphere approximation
(ASA) for one Nb-substituting Fe in the 2$\times$2$\times$2 supercell
of KNbO$_3$, without any oxygen vacancies\cite{icdim}, has shown
that the ``breathing'' relaxation of the O$_6$ octahedron is
$\sim$0.05{\AA} outwards. The shell model calculation for the same
geometry (also in Ref.~\onlinecite{icdim}) as well predicts
the outward expansion of O$_6$ of comparable magnitude.

The deficiencies of these previous calculations are that
the shell model depends on the empirical interaction parameters
fitted into it, and normally discards the anharmonic effects,
and the LMTO-ASA calculation is normally too crude for reliable
estimates of lattice relaxation. Moreover, no data on the
electronic structure of impurity was published in
Ref.~\onlinecite{icdim}. The aim of our present study is
to provide this information, concentrating specifically
on the treatment of different charge configurations within
the localized $d$-shell. In particular, we compare the treatment
within the local density approximation (LDA) with the so-called
LDA+$U$ approach. The calculations are done by a LMTO-ASA method
for undistorted lattice, without introducing oxygen vacancies.
The accurate analysis of lattice relaxation, with and without
vacancies, needs the use of a full-potential (FP, i.e.,
with the potential of general shape). Such FP-LMTO calculations
are in progress and will be reported elsewhere.

\section{Calculation method and setup}
\label{sec:setup}

The calculations have been done with the LMTO-ASA method\cite{oka},
in part using the TB-LMTO code of the O.~K.~Andersen group
in Stuttgart\cite{tblmto}. The LDA+$U$ calculation scheme
(discussed below) has been implemented in another LMTO-ASA code
by V.~I.~Anisimov. In all calculations, we considered the cubic
high-temperature phase of KNbO$_3$ with lattice constant 4.015{\AA};
atomic sphere radii were 1.639{\AA} for Fe and Nb, 1.964{\AA} for K
and 1.050{\AA} for O. Such choice of radii produces the band structure
in reasonable agreement with that from FP-LMTO calculations\cite{ktn3}.
According to our experience, the use of additional empty spheres
in the interstitials of the perovskite structure does not
bring any improvement in the LMTO-ASA calculation.
In order to account for the lifting of orbital degeneracy
at the Fe site, the minimal point group symmetry, including only inversion,
was assumed in the calculations.

Although KNbO$_3$ is not a typically ionic compound, and, moreover,
the covalency of the Nb--O bond is quite important for giving rise
to ferroelectric instability, it is useful to refer to nominal
ionic charges in order to understand why the problem of charge
compensation occurs, when one considers the Fe impurity.
Fe is either Fe$^{2+}$ or Fe$^{3+}$ in oxides; the substitution
of (nominally) Nb$^{5+}$ with Fe$^{3+}$ leaves 2 extra electrons
in the system, which can be removed along with one O$^{2-}$ ion,
thus restoring the neutrality. This is one possible mechanism
of local charge compensation; another, in principle, possible one
is that two Fe$^{3+}$ substitute Nb$^{5+}$ and K$^+$.
In terms of neutral atoms building the crystal, 30 electrons
are provided in total in the valence band per unit cell of KNbO$_3$
(including K$3p$, which are treated as valence band states).
If Fe atom with its 8 electrons substitutes Nb, only 5 electrons
can be absorbed by the valence band, leaving Fe nominally
in the $3d^3$ configuration. This is possible, but expectedly
highly instable. In order to restore a typical for Fe
$3d^5$ or $3d^6$ configuration, one should either provide
extra electrons, or remove one oxygen atom, thus reducing
the capacity of the valence band.
There are essentially only two ways to account for a charge
compensation in the calculation -- one either specifies explicitly
the configuration of impurities/vacancies that provides such
compensation, or adds extra electrons to the system,
implying that their donors are in some distant parts of crystal
and do not affect the local electronic structure.
The second way is probably technically easier, because
one needs only to search for the Fermi energy corresponding
to the specified number of extra electrons in each iteration.
In order to keep the supercell neutral, a compensating positive charge
is added in the background.

\begin{figure}
\epsfxsize=17.0cm\centerline{\epsffile{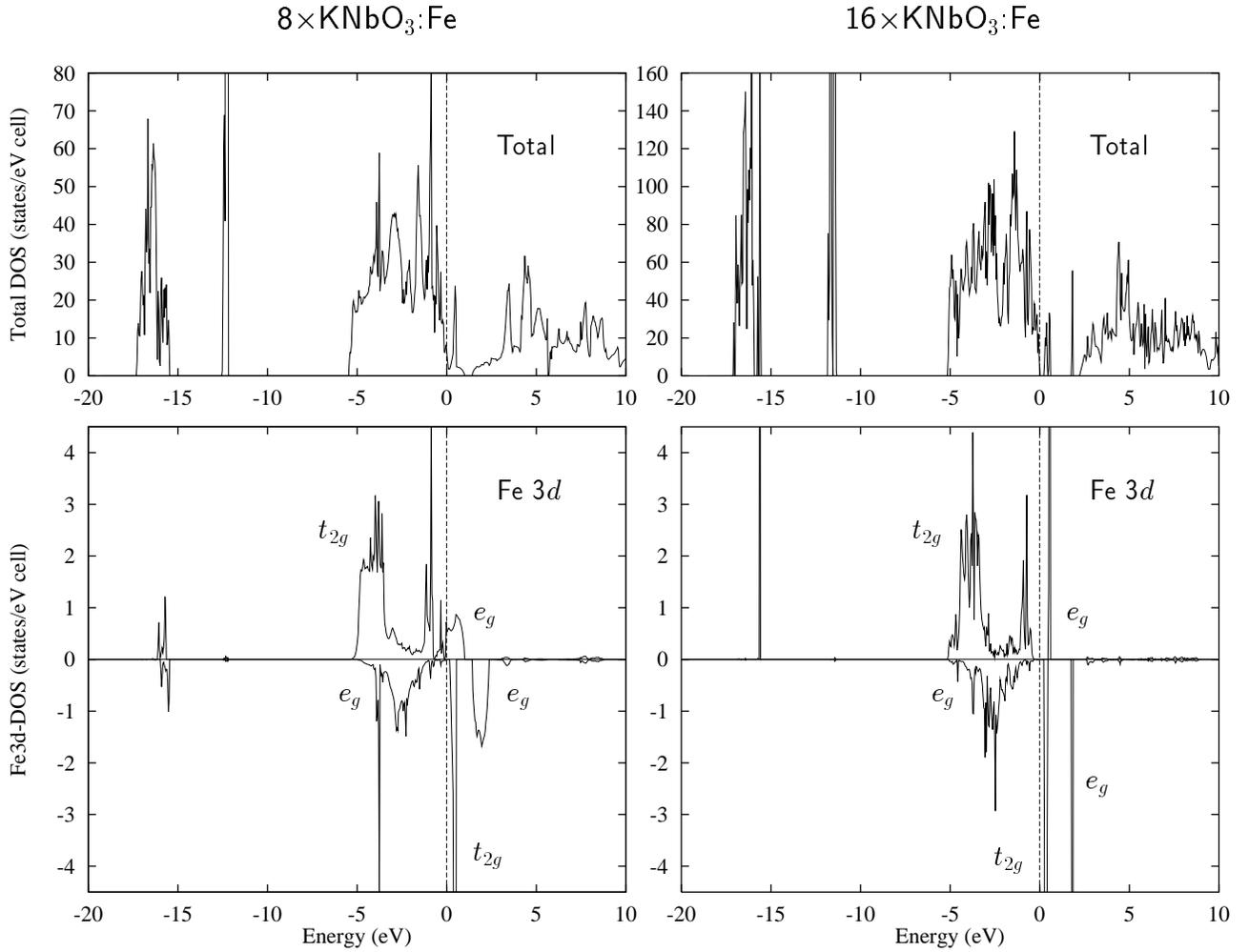}}
\caption{\protect\rule{0mm}{10mm}
Total and Fe$3d$-partial DOS as calculated
for 40-at. and 80-at. supercells of KNbO$_3$ with
single substitutional Fe impurity.
}
\label{fig:dos}
\end{figure}

Our initial calculations did not consider the charge
compensation in any way, simply treating the neutral
supercell with one Nb atom substituted with Fe.
As a smallest possible supercell for the modeling of
isolated impurity, we considered the perovskite cell doubled
in all three directions, i.e., including 40 atoms in total.
The density of states (DOS) resulting from this calculation
reveals the hybridization
of Fe$3d$ states with those of Nb and O over the whole
width of the valence band (see Fig.\ref{fig:dos}).
Moreover, several groups of localized impurity states
are split off to form flat bands just above the
Fermi level. Majority-spin $t_{2g}$ states are completely
filled, whereas all other $d$-orbitals are roughly
half-filled. Magnetic moment at Fe is 1.69 $\mu_{\mbox{\small B}}$
and exclusively related to the $3d$ shell.

It may be noted that the $t_{2g}$ states remain relatively
more localized than $e_g$, as should be expected already
from fact that the spatial distribution of the former dominates
along the line directed towards the interstitial between adjacent
K atoms. On the contrary, $e_g$ states directly overlap
with the $p$ shells of nearest oxygen atoms. This hybridization
mediates the interaction between the $e_g$-impurity levels
(for both spin directions) at adjacent Fe atoms which are situated only
two lattice spacings apart. As a result, the $e_g$-related
subbands are smeared so strongly that the band gap almost
disappears. We assume that the 8$\times$KNbO$_3$-supercell
is too small for a realistic modeling of isolated
Fe impurities.

The doubling of this supercell with the translation vectors
(022), (202), (220) results in the $\sqrt{2}$ times increased
distance between Fe atoms. The broadening of the $e_g$ states
is then sufficiently decreased to form completely split-off
discrete levels in the band gap. The system remains non-metallic,
in contrast to the 40-atom supercell where a small DOS
at the Fermi level was present. Therefore, this larger
80-atom supercell seems already to be appropriate for the
treatment of impurity problem. It should be noted however
that the charge and magnetic moment at the Fe site are almost
identical in the calculation with both supercells.

The number of electrons and magnetic moments within atomic spheres of
Fe and its nearest neighbors of different types are shown
in Table~\ref{tab:charge} for several numbers of extra electrons.
With no extra electrons, the charge configuration of Fe
is of course far from the nominal $3d^3$ as mentioned
above in reference to a non-charge compensated case. The reason
is, the electrons of the valence band which have been in perfect
crystal localized at the Nb site, should be, roughly speaking,
added to this number. Since KNbO$_3$ is not so strongly ionic,
and because of our choice of atomic sphere radii, the electron numbers
in Table~\ref{tab:charge} are close to those of neutral atoms.
Magnetic moment, induced by that of the impurity, primarily
resides on the first oxygen sphere.

\section{LDA+$U$ treatment}
\label{sec:lda+u}

\begin{figure}
\epsfxsize=14.0cm\centerline{\epsffile{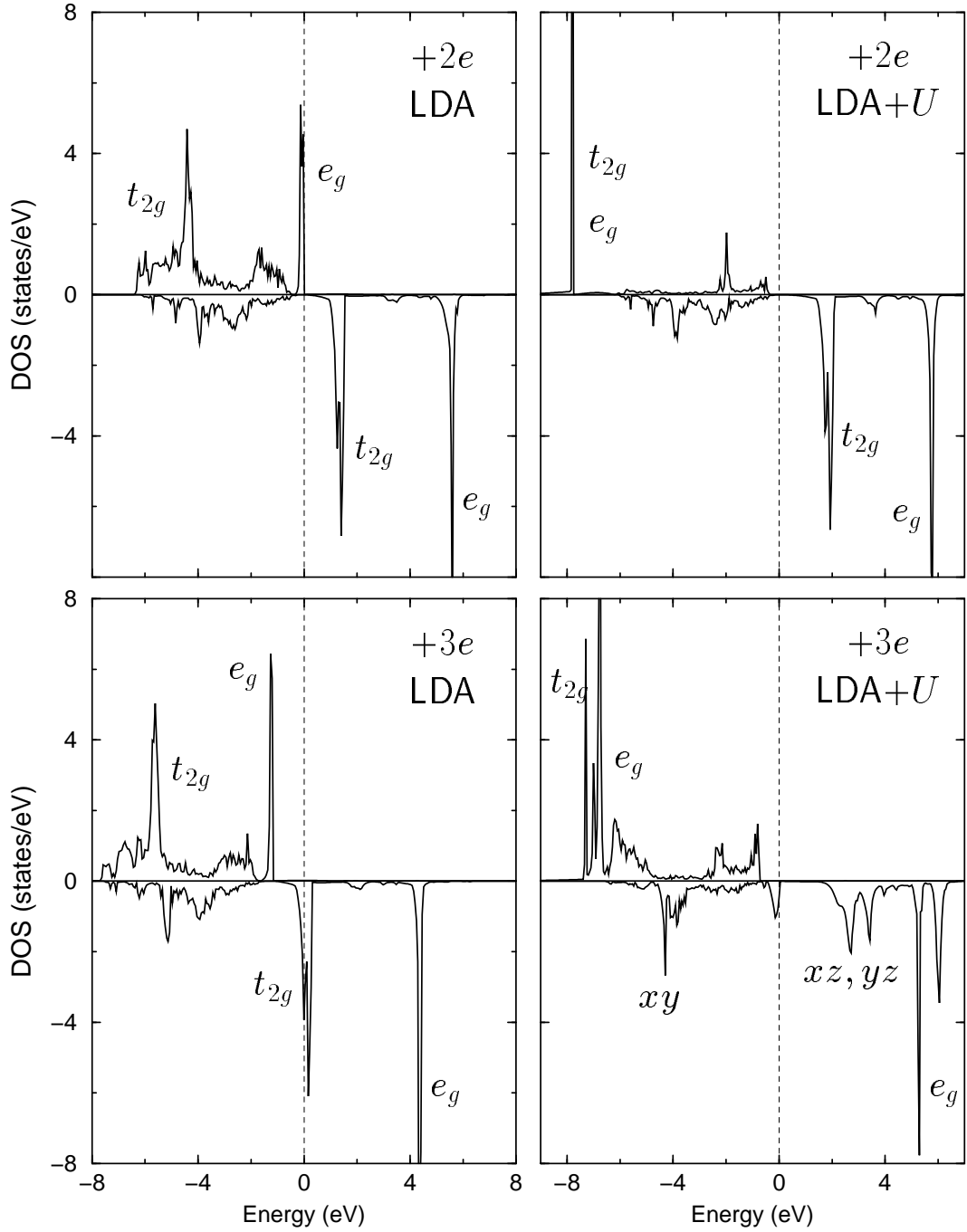}}
\caption{\protect\rule{0mm}{10mm}
Local $3d$-DOS for charge configurations with
2 and 3 addition electrons per 80-at. supercell
of KNbO$_3$:Fe, as calculated in the LDA and
LDA+$U$.
}
\label{fig:dos+u}
\end{figure}

In a similar supercell calculation for Fe in MgO \cite{mgo},
it was found that the LDA calculation results in a physically
wrong stable solution with zero magnetic moment, irrespectively
of the supercell size. In the present system, Fe impurity
has similar nearest neighborhood of oxygen octahedron, and
the crystal field splits the $3d$-states into lower $t_{2g}$
and upper $e_g$ states in a similar way, but the solution
is magnetic and stable. Apparently, the difference from the
MgO:Fe case is in the missing nearest neighbors in the [110]
direction in the perovskite lattice. In MgO, non-magnetic Mg
atoms interact strongly with Fe $t_{2g}$ states and force
the occupation numbers in both spin channels to become equal.
This mechanism works in the LDA, because the potential
acting on the $t_{2g}$ states is not dependent on their
occupation, and dominates over the intraatomic exchange.

Although no obvious problem of such kind comes out for
for Fe at the Nb site in KNbO$_3$, one should keep in mind that
the treatment of localized $d$ states in the LDA may in
principle be erroneous. A practically feasible way to deal
with this problem is the LDA+$U$ approach, proposed by
Anisimov {\it et al.}\cite{lda+u} and tested by now for many systems,
including Fe oxides\cite{fe3o4,feo} and some perovskites\cite{perov}.
The essence of the method is that for the states which are
{\it ad hoc} declared as localized, the total-energy functional
as provided by the LDA and casted in the form dependent on the
total number of localized electrons is corrected depending
on the actual occupation numbers of different localized orbitals.
As a result, the potentials acting on localized states
become occupation dependent, and the one-electron energies
get accordingly corrected, thus lifting orbital degeneracies
when physically motivated. Effective Coulomb interaction $U$
and effective intraatomic exchange $J$ enter as external parameters
in a self-consistent cycle of LDA+$U$ calculation. These parameters
can be determined from first principles
(see, e.g., Ref.~\onlinecite{param}), once the orbitals which
should be treated as localized are selected. However, $U$ and $J$
are of essentially intraatomic character and may be
largely transferred within many systems including the same
chemical constituent with {\it a priori} localized states.

Solovyev {\it et al.}\cite{lamo3} thoroughly studied the applicability
of LDA+$U$ in comparison with LDA to La$M$O$_3$ ($M$=Ti--Cu) perovskite
systems, where a similar problem of treating more localized
$t_{2g}$ and somehow less localized $e_g$ states of a $3d$ constituent
occurs. As apparently the most reasonable option (among others tested),
the $t_{2g}$ states have been singled out as localized ones,
to be treated according the prescription of the LDA+$U$ approach,
while $e_g$ have been attributed to common band states well
describable within the LDA. This segregation may prove useful
for the Fe impurity as well. But as a first try, we treated
all Fe$3d$ states as localized.  Estimates of $U$ for Fe
based on somehow different criteria vary from $\sim$9 eV\cite{lamo3}
to 5.1 eV\cite{feo}; we used in our calculation $U$=6.8 eV as estimated
in Ref.~\onlinecite{gunnar}, and $J$=0.89 eV.
We did the LDA+$U$ calculations only for the systems with
added 2 and 3 electrons, that correspond to experimentally expected
(nominally) Fe$^{3+}$ and Fe$^{2+}$ configurations, correspondingly.
The related lines of data in Table~\ref{tab:charge} are labeled (+$U$).

\section{Overview of Results}
\label{sec:results}

\begin{table}
\caption{
Charges $Q$ and magnetic moments $M$ within atomic spheres
of Fe impurity and its several neighboring atoms as calculated
for different number of added electrons
}
\label{tab:charge}
\begin{tabular}{cdddddddddddd}
                               &
 \multicolumn{2}{c}{Fe}        &
 \multicolumn{2}{c}{O$_{xy}$}  &
 \multicolumn{2}{c}{O$_{z}$}   &
 \multicolumn{2}{c}{Nb$_{xy}$} &
 \multicolumn{2}{c}{Nb$_{z}$}  &
 \multicolumn{2}{c}{K}         \\
\cline{2-13}
\raisebox{2.0ex}[0pt]{~~~extra $e$~~~} &
            $Q$  &  $M$ & $Q$  & $M$  & $Q$  & $M$  &
	    $Q$  & $M$  & $Q$  &   $M$  & $Q$  & $M$  \\
\hline
 0        & 8.78 & 1.69 & 5.88 & 0.24 & 5.90 & 0.27 &
	    4.96 & 0.01 & 5.04 &$-$0.01 & 6.98 & 0.00 \\
 1        & 8.79 & 1.68 & 5.90 & 0.19 & 5.91 & 0.24 &
	    4.98 & 0.01 & 5.05 &   0.00 & 6.99 & 0.00 \\
 2        & 8.93 & 3.13 & 5.90 & 0.43 & 5.90 & 0.53 &
	    5.02 & 0.02 & 5.09 &   0.01 & 7.00 & 0.00 \\
 2 (+$U$) & 8.85 & 3.31 & 5.92 & 0.41 & 5.91 & 0.51 &
	    5.01 & 0.02 & 5.09 &   0.01 & 7.01 & 0.00 \\
 3        & 8.92 & 3.13 & 5.93 & 0.33 & 5.93 & 0.42 &
	    5.03 & 0.04 & 5.10 &   0.03 & 7.02 & 0.00 \\
 3 (+$U$) & 8.88 & 3.20 & 5.93 & 0.34 & 5.93 & 0.45 &
	    5.05 & 0.02 & 5.11 &   0.00 & 7.02 & 0.00 \\
\end{tabular}
\end{table}

As is seen from Table~\ref{tab:charge}, the number of electrons
on all constituents, at the exception of Fe, grows very
smoothly with the extra background charge. As the first electron
is added, nothing happens at the Fe site, and the extra electron
goes completely into the valence band, resulting only in a slight
shift of the Fermi energy but not visibly affecting the DOS.
With the second extra electron, the Fermi level finally
crosses the narrow majority-spin $e_g$ subband (Fig.~\ref{fig:dos+u}).
This increases the magnetic moment considerably
(by less that 2$\mu_{\mbox{\small B}}$, however, because the majority-spin
$e_g$ states also contribute somehow to the already occupied valence band).
The distribution of the impurity states in the minority-spin
channel is affected only quantitatively. The third extra electron
begins to populate the minority-spin $t_{2g}$ subband,
but most part of it goes into the valence band, where the
charge at all sites almost uniformly increases.
It is noteworthy that the magnetic moment at O neighbors
dramatically increases with the addition of the second electron,
but comes back (and the charge increases, instead) with the
addition of the third one. Since the states at the top
of the valence band are of exclusively O$2p$ character
(see, e.g., Ref.~\onlinecite{ktn3}), the drift of the
$e_g$-subband below the Fermi level and into the valence band
affects the majority-spin O$2p$ states most directly.
The crossing of the Fermi level by minority-spin $t_{2g}$ subband
contributes to the minority-spin O$2p$ subband and reduces
the magnetic moment on oxygen atoms, increasing at the same time
their occupation. At this point, the majority-spin $e_g$ subband
enters the region of noticeable Nb contribution in
the valence band and produces the peak of magnetic moment
on Nb sites.

Comparing the results of LDA and LDA+$U$ calculations
(Fig.~\ref{fig:dos+u}), one should keep in mind that the
overall effect of the latter is the lowering the energies
of occupied states and the upward shift of vacant ones.
This is exactly what happens in the configuration with 2
extra electrons. Since essentially all majority-spin states
are already occupied in this configuration and all
minority-spin states empty, the inclusion of the $U$-correction
has a negligible effect on all integral properties
(both charges and moments). However, the exact position
of the $t_{2g}$-impurity state is changed, that may give
rise to the change in the optical absorption.
The situation in the configuration with 3 extra electrons
is completely different. Since here the minority-spin
$t_{2g}$ state is partly occupied in the LDA, in the
LDA+$U$ treatment the potentials acting on the
$xy$, $xz$ and $yz$ components of it become different,
thus lifting the orbital degeneracy of these states
and distorting the local DOS considerably.

Summarizing, our study of the electronic structure
of Fe impurity in the Nb site of KNbO$_3$ with the analysis of different
charge compensation has shown, that only two considerably
different configurations of impurity occur. The first one,
with the magnetic moment $\sim$1.7 $\mu_{\mbox{\small B}}$,
corresponds to the substitutional impurity without compensation --
the situation that probably is not very common in reality.
This configuration, however, survives under addition of one
extra electron per impurity. Two extra electrons --
or, equivalently, Fe impurity combined with a distant
oxygen vacancy -- induces a transition into the high-spin state
with the magnetic moment $\sim$3.1 $\mu_{\mbox{\small B}}$,
with the minority-spin $t_{2g}$ level in the band gap.
The charge and magnetic moment
of this configuration remain intact with the addition of
third extra electron, but the exact position of the level
in the gap (probably, its splitting as well) may be affected.
The last two configurations with 2 or 3 extra electrons
correspond to the most practically relevant impurity configurations,
referred to as Fe$^{3+}$ and Fe$^{2+}$.
A more precise study of their energetics is possible with the
use of a full-potential calculation scheme and simultaneous
analysis of lattice relaxation around impurity and, optionally,
oxygen vacancy.

\acknowledgements

The authors are grateful to V.~Anisimov for his assistance and
providing the code for LDA+$U$ calculations, and to M.~Korotin
for useful discussions.
Financial support of the Deutsche Forschungsgemeinschaft (SFB~225)
is gratefully acknowledged. A.I.P. appreciates the support
of the NATO International Scientific Exchange Program
(Grant HTECH.LG 940861) and the hospitality of the University
of Osnabr\"uck during his stay there.


\begin{references}
\bibitem{review} {\it Photorefractive Materials and Their Applications I:
		 Fundamental Phenomena},
		 ed. P.~G\"unter and J.-P.~Huignard
		 (Springer-Verlag, Berlin Heidelberg, 1988), pp. 131-66.
\bibitem{buse95} K.~Buse and E.~Kr\"atzig,
		 {\it Appl.~Phys.~B} {\bf 61}, 27 (1995).
\bibitem{medra}  C.~Medrano, E.~Voit, P.~Amrhein, and P.~G\"unter,
		 {\it J.~Appl.~Phys.} {\bf 64}, 4668 (1988).
\bibitem{siegel} E.~Siegel, W.~Urban, K.~A.~Mueller, and E.~Wiesendanger,
		 {\it Phys.~Lett.} {\bf 53}A, 415 (1975);
		 E.~Siegel,
		 {\it Ferroelectrics} {\bf 13}, 385 (1976).
\bibitem{donner} E.~Possenriede, O.~F.~Schirmer, H.~J.~Donnerberg,
		 and B.~Hellermann,
		 {\it J.~Phys.:~Cond.~Matter} {\bf 1}, 7267 (1989).
\bibitem{exner}  M.~Exner,
		 {\it Computer-Simulation von extrinsischen und intrinsischen
		 Defekten in Kaliumtantalat- und Kaliumniobat-Kristallen}
		 (Shaker, Aachen, 1994).
\bibitem{lewis}  G.~V.~Lewis and C.~R.~A.~Catlow,
		 {\it J.~Phys.~Chem.~Solids} {\bf 47}, 89 (1986).
\bibitem{beck}   O.~Beck, D.~Kollewe, A.~Kling, W.~Heiland, and F.~Hesse,
		 {\it Nucl.~Instr.~Meth.~Res.~B} {\bf 85}, 474 (1994).
\bibitem{icdim}  H.~Donnerberg, M.~Exner, T.~Neumann, G.~Borstel,
		 O.~F.~Schirmer, R.~H.~Bartram, and C.~R.~A.~Catlow,
		 {\it Defects in Insulating Materials --
		 Proceedings of the XII Inernat. Conference on,
		 Vol. 2} (World Scientific, Singapore, New Jersey, 1993),
		 pp. 1178-80.
\bibitem{oka}    O.~K.~Andersen,
		 {\it Phys.~Rev.~B} {\bf 12}, 3060 (1975).
\bibitem{tblmto} O.~K.~Andersen and O.~Jepsen,
		 {\it Phys.~Rev. Lett.} {\bf 53}, 2571 (1984);
                 O.~K.~Andersen, Z.~Pawlowska, and O.~Jepsen,
	         {\it Phys. Rev. B} {\bf 34}, 5253 (1986).
\bibitem{ktn3}   A.~V.~Postnikov, T.~Neumann, G.~Borstel, and
		 M.~Methfessel,
	         {\it Phys.~Rev.~B} {\bf 48}, 5910 (1993).
\bibitem{mgo}    M.~A.~Korotin, A.~V.~Postnikov, T.~Neumann,
		 G.~Borstel, V.~I.~Anisimov, and M.~Methfessel,
		 {\it Phys.~Rev.~B} {\bf 49}, 6548 (1994).
\bibitem{lda+u}  V.~I.~Anisimov, J.~Zaanen, and O.~K.~Andersen,
		 {\it Phys.~Rev.~B} {\bf 44}, 943 (1991);
		 A.~I.~Liechtensiein, J.~Zaanen, and V.~I.~Anisimov,
		 {\it Phys.~Rev.~B} {\bf 52}, R5467 (1995);
		 V.~I.~Anisimov, F.~Aryasetiawan and A.~I.~Lichtenstein,
		 {\it J.~Phys.:~Cond.~Matter} {\bf 9}, 767 (1997).
\bibitem{fe3o4}  V.~I.~Anisimov, I.~S.~Elfimov, N.~Hamada, and K.~Terakura,
		 {\it Phys.~Rev.~B} {\bf 54}, 4387 (1996).
\bibitem{feo}    I.~I.~Mazin and V.~I.~Anisimov,
		 {\it cond-mat/9610147}.
\bibitem{perov}  M.~A.~Korotin, S.~Yu.~Ezhov, I.~V.~Solovyev,
		 V.~I.~Anisimov, D.~I.~Khomskii, and G.~Sawatzky,
		 {\it Phys.~Rev.~B} {\bf 54}, 5309 (1996);
		 V.~I.~Anisimov, I.~S.~Elfimov, M.~A.~Korotin, and K.~Terakura,
		 {\it cond-mat/9609158}.
\bibitem{param}  I.~V.~Solovyev, P.~H.~Dederichs, and V.~I.~Anisimov,
		 {\it Phys.~Rev.~B} {\bf 50}, 16861 (1994).
\bibitem{gunnar} V.~I.~Anisimov and O.~Gunnarsson,
		 {\it Phys.~Rev.~B} {\bf 43}, 7570 (1991).
\bibitem{lamo3}  I.~Solovyev, N.~Hamada, and K.~Terakura,
		 {\it Phys.~Rev.~B} {\bf 53}, 7158 (1996).
\end{references}
\end{document}